\documentclass[aps,prl,twocolumn,amsmath,amssymb,showpacs]{revtex4}
\usepackage{graphicx}
\usepackage{dcolumn}
\usepackage{bm}
\begin{document}

\title{Electric Spaser in the Extreme Quantum Limit}

\author{Dabing Li}
\affiliation{State Key Laboratory of Luminescence and Applications, Changchun Institute of Optics, Fine Mechanics and Physics, Chinese Academy of Sciences,
Changchun 130033, China
}
\author{Mark I. Stockman}
\affiliation{
Department of Physics and Astronomy, Georgia State University, Atlanta, Georgia 30303, USA
}

\date{\today}
\begin{abstract}
We consider theoretically the spaser excited electrically via a nanowire with ballistic quantum conductance. We show that in the extreme quantum regime, i.e., for a single conductance-quantum nanowire, the spaser with the core made of common plasmonic metals, such as silver and gold, is fundamentally possible. For ballistic nanowires with multiple-quanta or non-quantized conductance, the performance of the spaser is enhanced in comparison with  the extreme quantum limit. The electrically-pumped spaser is promising as an optical source, nanoamplifier, and digital logic device for optoelectronic information processing with speed $\sim100 ~\mathrm{GHz}$ to $\sim100 ~\mathrm{THz}$.
\end{abstract}

\maketitle 

Active  or gain nanoplasmonics was introduced  \cite{Bergman_Stockman:2003_PRL_spaser} by spaser (surface plasmon amplification by stimulated emission of radiation). The spaser is a nanoscale quantum generator and ultrafast nanoamplifier of coherent localized optical fields \cite{Bergman_Stockman:2003_PRL_spaser, Stockman_Nat_Phot_2008_Spasers_Explained, Stockman_JOPT_2010_Spaser_Nanoamplifier, Sorger_Zhang_Science_2011_Spasers, Stockman_Opt_Expres_2011_Nanoplasmonics_Review}. The spaser is a nanosystem  constituted by a plasmonic metal and a gain medium. The spaser is based on compensation of optical losses in metals by gain in the active medium (nanoshell) overlapping with the surface plasmon (SP) eigenmodes of the metal plasmonic nanosystem. There are many experimentally observed and investigated spasers where the gain medium consisted of dye molecules \cite{Noginov_et_al_Nature_2009_Spaser_Observation, Odom_et_al_Nano_Lett_2012_Bowtie_Spaser_Array, Noginov_et_al_J_Opt_2012_SPP_Spasing_on_Smooth_and_Rough_Surfaces}, unstructured semiconductor nanostructures and nanoparticles \cite{Hill_et_al_2007_Nat_Phot_2007_Nanolasers,
Oulton_Sorger_Zentgraf_Ma_Gladden_Dai_Bartal_Zhang_Nature_2009_Nanolaser, Hill_et_al_Opt_Expr_2009_Polaritonic_Nanolaser,
 Zhang_et_al_Nature_Materials_2010_Spaser,  Hill_et_al_Opt_Expr_2011_DFB_SPP_Spaser, Gwo_et_al_NanoLett_2011_Plasmonic_Green_Spaser_GaN, IEEE_J_Quant_Electron_2011_SP_Nanosisk_Lasers,  Ning_et_al_PRB_2012_CW_Lasing_in_Deep_Subwavelengh_Metal_Nanocavities,
 Zhang_et_al_Nano_Lett_2012_Muliticolor_Spaser, Gwo_et_al_Science_2012_Spaser}, or 
 quantum-confined semiconductor heterostructures: quantum dots (QDs) \cite{Zheludev_et_al_Nat_Phot_2008_Lasing_Spaser, Plum_Fedotov_Kuo_Tsai_Zheludev_Opt_Expr_2009_Toward_Lasing_Spaser}, quantum wires (QWs), or quantum wells \cite{Long_et_al_Opt_Expr_2011_Spaser_1.5micron_InGaAs}. 

Classified by mode confinement, there are the spasers with one-dimensional  \cite{Hill_et_al_2007_Nat_Phot_2007_Nanolasers, Hill_et_al_Opt_Expr_2009_Polaritonic_Nanolaser,  Hill_et_al_Opt_Expr_2011_DFB_SPP_Spaser}, two-dimensional \cite{Oulton_Sorger_Zentgraf_Ma_Gladden_Dai_Bartal_Zhang_Nature_2009_Nanolaser}, or three-dimensional (3d) confinement \cite{Noginov_et_al_Nature_2009_Spaser_Observation, Odom_et_al_Nano_Lett_2012_Bowtie_Spaser_Array, Gwo_et_al_Science_2012_Spaser}.
The spasers can also be classified by the spasing-eigenmode type, which can be either localized surface plasmons (SPs)  \cite{Noginov_et_al_Nature_2009_Spaser_Observation, Gwo_et_al_NanoLett_2011_Plasmonic_Green_Spaser_GaN, Odom_et_al_Nano_Lett_2012_Bowtie_Spaser_Array, Gwo_et_al_Science_2012_Spaser}, or surface plasmon polaritons (SPPs) \cite{Gramotnev_Bozhevolnyi_Nat_Phot_2010_Review_SPPs_Concentration, Han_Bozhevolnyi_Rep_Progr_Phys_2013_SPP_Review} as in the rest of the cases. Among the observed spasers, most are with optical pumping,
including all the spasers with the strong 3d confinement \cite{Noginov_et_al_Nature_2009_Spaser_Observation, Gwo_et_al_NanoLett_2011_Plasmonic_Green_Spaser_GaN, Odom_et_al_Nano_Lett_2012_Bowtie_Spaser_Array, Gwo_et_al_Science_2012_Spaser}.
Only few SPP spasers, whose confinement and losses are not strong, are with electric pumping \cite{Hill_et_al_2007_Nat_Phot_2007_Nanolasers, Hill_et_al_Opt_Expr_2009_Polaritonic_Nanolaser,  Hill_et_al_Opt_Expr_2011_DFB_SPP_Spaser}. 

\begin{figure}
\begin{center}\includegraphics[width=0.29\textwidth]{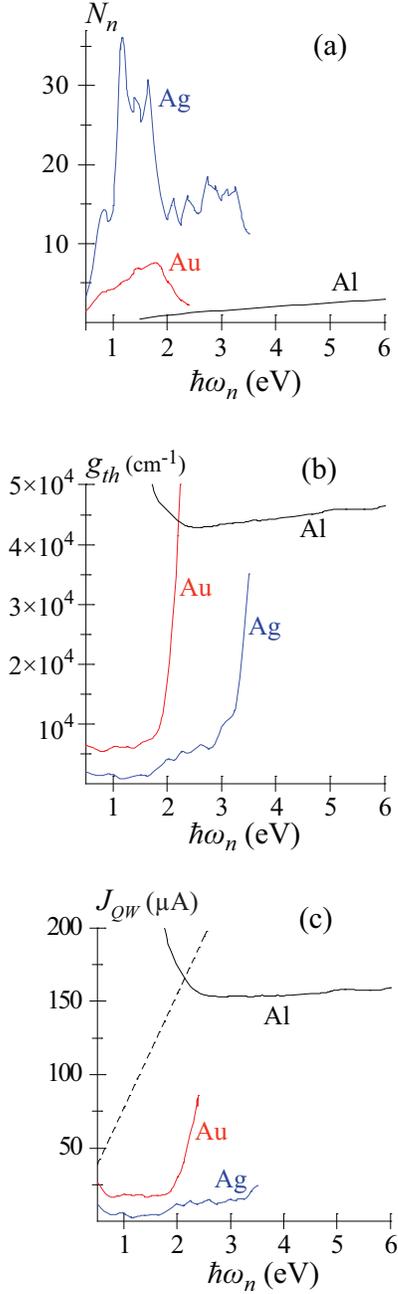}\end{center}
\caption{For three plasmonic metals, Ag, Au, and Al, dependencies of working parameters of the CW-regime spaser on spasing frequency $\omega_n$ expressed as the corresponding SP energy $\hbar\omega$. (a) Mean number of SPs per spasing mode $N_n$ in the extreme quantum limit: computed from Eq.\ (\ref{NQ}) for a single conduction quantum channel, $N_c=1$. (b) Threshold gain $g_{th}$ required for spasing as computed from Eq.\ (\ref{g_th}). (c) The nanowire current required for developed spasing with a single SP quantum per the spasing mode for the three plasmonic metals as indicated (solid lines). The current of the single-channel QW under the minimum required potential difference is shown by the dashed line. The developed spasing is possible for the spectral regions where  the corresponding $J_{QW}$ curves are below this line.
}
\label{Spasing_Conditions}
\end{figure}

There have been doubts expressed in the literature regarding viability of the nanospaser with the strong 3d confinement \cite{Khurgin_Sun_APL_2011_Scaling_of_Losses_in_Plasmonics}, especially with electrical pumping \cite{Khurgin_Sun_opt_Expr_2012_Impossibility_Electrically_Pumped_Spaser}. In a direct contrast,  the possibility of the optically pumped strongly 3d-confined spaser has been both theoretically shown \cite{Bergman_Stockman:2003_PRL_spaser, Stockman_JOPT_2010_Spaser_Nanoamplifier, Stockman_PRL_2011_Loss_Compensation, Stockman_Phil_Trans_R_Soc_A_2011_Spasing_Loss_Compensation} and experimentally demonstrated \cite{Noginov_et_al_Nature_2009_Spaser_Observation, Gwo_et_al_NanoLett_2011_Plasmonic_Green_Spaser_GaN, Odom_et_al_Nano_Lett_2012_Bowtie_Spaser_Array, Gwo_et_al_Science_2012_Spaser}. Here we theoretically establish that the electrically-pumped spaser is fundamentally possible. 
A principal difference of our theory from the previous works \cite{Khurgin_Sun_APL_2011_Scaling_of_Losses_in_Plasmonics, Khurgin_Sun_opt_Expr_2012_Impossibility_Electrically_Pumped_Spaser} is that we consider ballistic, quantum electron transport instead of classical, dissipative one. Another important difference is that the contradicting theoretical work  \cite{Khurgin_Sun_APL_2011_Scaling_of_Losses_in_Plasmonics, Khurgin_Sun_opt_Expr_2012_Impossibility_Electrically_Pumped_Spaser} on spasers (sometimes also called plasmonic nanolasers) has ignored distinction between the threshold condition of spasing and the condition of developed spasing with $N_n\sim1$ quanta (SPs) per generating mode. While for  macroscopic lasers with enormously large number of photons this distinction is not significant, for spasers, where $N_n\sim 1$, it is very  important as we show below in conjunction with Fig.\ \ref{Spasing_Conditions}.

One of the most important problems in nanoplasmonics, both fundamentally and from the application standpoint, is the development of electrically-pumped 3d-confined nanospaser, which can serve as a versatile nanoscopic source of optical near-field energy or ultrafast nanoscale amplifier which has the same form form factor as the field-effect transistor ($\lesssim 30$ nm linear dimensions) but is $\sim 10^3$ times faster \cite{Stockman_JOPT_2010_Spaser_Nanoamplifier} with prospects of the large-scale integration in petahertz-bandwidth systems. Such a spaser has not yet been observed. This Letter demonstrates that the electric spaser in the extreme quantum regime, i.e., pumped under the minimum required voltage via a quantum wire with a single quantized conduction channel, is fundamentally possible.

Our goal is to formulate a condition of spaser operation in the extreme quantum limit where it is only limited by laws of quantum mechanics and expressed through fundamental constants. The corresponding model is comprised by a single-mode spaser with electric excitation via a QW (or a quantum point contact) with $N_c$ open conduction channels. The conductivity of such a QW is 
 \cite{ Sharvin_JETP_1965_Quantum_Point_Contact, Landauer_1985_Conduction_Quantum, Foxon_et_al_PRL_1988_Quantized_Resistance, Wharam_et_al_J_Phys_C_1988_Quantized_Resistance} 
\begin{equation}
G=N_c G_0~,~~~G_0=\frac{e^2}{\pi\hbar}~,
\label{G_0}
\end{equation}
where $G_0$ is the conduction quantum, and $e$ is elementary charge; $N_c=1$ in the extreme quantum limit. 
Such a QW may be a ballistic semiconductor channel \cite{Foxon_et_al_PRL_1988_Quantized_Resistance, Wharam_et_al_J_Phys_C_1988_Quantized_Resistance},  a metal point contact \cite{ Sharvin_JETP_1965_Quantum_Point_Contact, Gai_et_al_PRB_1996_Quantized_Metal_Point_Contacts_at_Room_T, Houten_Beenakker_Phys_Today_1996_Quantum_Point_Contact},  carbon nanotube ($N_c=2$) \cite{Dresselhaus_et_al_2001_Carbon_Nanotubes}, or a graphene sheet \cite{deHeer_et_al_Science_2006_Confinement_in_Graphene, Novoselov_et_al_Nature_2012_Graphene_Review}.

In the same spirit of the extreme quantum limit, we assume perfect (hundred percent) quantum efficiency of the electron to SP energy conversion. This requires that the potential difference $U$ along the QW accelerates an electron to acquire just the energy of one plasmon, $U=\hbar\omega_n/e$, where $\omega_n$ is the spasing frequency.  In view of Eq.\ (\ref{G_0}), this results in the current along the QW, 
\begin{equation}
J_{QW}=N_c e\omega_n/\pi~.
\label{J_QW}
\end{equation}
We assume developed spasing where the stimulated emission into the spasing mode dominates over the spontaneous emission. As theory of the spaser has demonstrated \cite{Stockman_JOPT_2010_Spaser_Nanoamplifier}, due to the very strong feedback in the nanospaser, the developed spasing occurs already for $N_n\gtrsim 1$. Under such conditions, the above-described excitation process leads to accumulation of the mean number of SPs per spasing mode in the stationary (continuous wave, or CW) regime, 
\begin{equation}
N_n= \tau_n J_{QW} /e~, 
\label{N_n}
\end{equation}
where $\tau_n=1/(2\gamma_n)$ is the SP lifetime, $\gamma_n=\gamma_n(\omega_n)$ is the SP amplitude spontaneous decay rate \cite{Bergman_Stockman:2003_PRL_spaser, Stockman_Opt_Expres_2011_Nanoplasmonics_Review} at $\omega_n$,
\begin{equation}
\gamma_n={\mathrm{Im}\,\varepsilon_m(\omega_n)}\left/{\frac{\partial\mathrm{Re}\,\varepsilon_m(\omega_n)}{\partial\omega_n}}\right. ~,
\label{gamma_n}
\end{equation}
and $\varepsilon_m$ is the permittivity of the spaser plasmonic metal. 

Substituting Eq.\ (\ref{J_QW}) into Eq.\ (\ref{N_n}) we obtain a fundamental expression for the mean number of the SP quanta per the spasing mode for an spaser pumped electrically via a single-quantum-of-conduction channel,
\begin{equation}
N_n=N_c Q(\omega_n)/\pi~,~~~ Q(\omega_n)=\omega_n/(2\gamma_n)~,
\label{NQ}
\end{equation}
where $Q(\omega_n)$ is the SP quality factor at the spasing frequency in its standard definition \cite{Stockman_Opt_Expres_2011_Nanoplasmonics_Review}. Note that the expression for $N_n$  (\ref{NQ}) depends only on the spasing frequency and the permittivity of the plasmonic metal, which in turn, is determined by the fundamental constants ($e$, $\hbar$, and electron mass $m$) only. It does not depend explicitly on geometry of the spaser or properties of the spaser gain medium and the QW. 
As we show below (see the following paragraphs), the condition of the developed spasing, $N_n\gtrsim 1$, where $N_n$ is given by Eq.\ (\ref{NQ}), is fundamentally different from the spaser threshold condition. 

 In the extreme quantum regime of $N_c=1$, the mean number of SPs per spasing mode, $N_n$ (\ref{NQ}), is plotted in Fig.\ \ref{Spasing_Conditions} (a) for three plasmonic metals: silver (Ag), gold (Au), and aluminum (Al). The values of $\varepsilon_m$ of Ag and Au used are from Ref.\ \onlinecite{Johnson:1972_Silver} and of Al from Ref.\ \onlinecite{Palik_1985_Optical_Constants}.

Figure \ref{Spasing_Conditions} (a) clearly demonstrates that the electrically pumped spaser in the extreme quantum limit fundamentally can undergo the developed generation: the number of the SP quanta per spasing mode is $N_n=15-30$ for Ag in the near infrared (ir) and visible (vis) spectral regions, in the near-ir and red-yellow vis spectrum, $N_n=6-7$, while for Al $N_n=2-3$ in the ultraviolet spectral region. 

Compare the condition of the developed spasing, $N_n\ge 1$, with the condition \cite{Stockman_JOPT_2010_Spaser_Nanoamplifier}  for the spaser threshold, which is obtained for $N_n=0$ and infinitely strong pumping (full population inversion). For self-containment, we give below this threshold condition in the form of Ref.\ \onlinecite{Stockman_Opt_Expres_2011_Nanoplasmonics_Review},
\begin{equation}
g\ge g_{th}~,~~~ g_{th}=\frac{\omega_n}{c\sqrt{\varepsilon_d}}\frac{\mathrm{Re}\, s(\omega_n)}{1-\mathrm{Re}\, s(\omega_n)}\mathrm{Im}\, \varepsilon_m(\omega_n)~,
\label{g_th}
\end{equation}
where $g$ is the gain that an infinite gain medium made up of the spaser gain material should have, $c$ is speed of light, $\varepsilon_d$ is the permittivity of the  dielectric medium into which the spaser is embedded, and $s(\omega)=[1-\varepsilon_m(\omega)/\varepsilon_d]^{-1}$ is Bergman's spectral parameter \cite{Bergman_Stroud_1992}. The gain, $g$, of the gain medium is given by the standard expression,
\begin{equation}
g=\frac{4\pi\omega_n}{3c}\frac{\sqrt{\varepsilon_d} |\mathbf d_{12}|^2 n_c}{\hbar \Gamma_{12}}~,
\label{g}
\end{equation}
where $\mathbf d_{12}$ is the transition dipole matrix elements for the spasing transition in  the gain medium, $n_c$ is the density of this transitions (i.e., the density of the excitons or recombining electron-hole pairs in the gain medium), and $\Gamma_{12}$ is the spectral width of the spasing transition.

To illustrate  quantitative differences between the developed spasing and the spasing threshold, we plot in Fig.\ \ref{Spasing_Conditions} (b) threshold gain $g_{th}$ as a function of the spasing frequency expressed as energy $\hbar\omega_n$ of a SP quantum. Note that a realistic gain that one can obtain for dye molecules or direct band gap semiconductors is $g\sim 10^4 ~\mathrm{cm^{-1}}$. Thus, it is obvious from  Fig.\ \ref{Spasing_Conditions} (b) that one can build nanospasers with silver or gold as plasmonic metals, in agreement with  experimental demonstrations \cite{Noginov_et_al_Nature_2009_Spaser_Observation, Gwo_et_al_Science_2012_Spaser}, but an aluminum-based spaser would require $g_{th}\sim 10^5~\mathrm{cm^{-1}}$, which is very high, thus making spasing on aluminum problematic. At the same time,  the mean SP population numbers for Au and Al in the developed spasing regime illustrated in  Fig.\ \ref{Spasing_Conditions} (a) are on the same order, $N_n\sim 1-5$. 

Hence, the condition of developed spasing,  $N_n\gtrsim 1$, is fundamentally different from the  spasing threshold conditions, e.g.,  Eqs.\ (\ref{g_th})-(\ref{g}). The reason is that  the  expression (\ref{N_n}) for the mean number of quanta per a single generating mode is only valid for the case of the developed spasing, i.e., well above the spasing threshold when the stimulated emission dominates over the spontaneous one. This is evident already from the fact that the coherent SP population, $N_n$, in Eq.\  (\ref{N_n}) is proportional to the pumping current. It depends only on spasing frequency  $\omega_n$ and SP quality factor $Q(\omega_n)$, which, in turn, only depends on metal permittivity $\varepsilon_m$. Equation  (\ref{N_n}) does not depend whatsoever on the properties of the gain medium. 

In a sharp contrast, the threshold conditions (\ref{g_th})-(\ref{g}) are those in the absence of the coherent SP population, i.e., for $N_n=0$. Physically, they mean that the unphased system of the nanoplasmonic core and the gain shell at the threshold of spasing becomes unstable. Above the threshold, this instability is resolved by a non-equilibrium phase transition with a spontaneous break down of symmetry: a coherent state of the spaser emerges with a finite mean SP population, $N_n$  whose phase is established completely randomly but is sustained afterwards. Note that the threshold conditions Eqs.\ (\ref{g_th})-(\ref{g}) impose  requirements to the gain medium and do contain its microscopic parameters $\mathbf d_{12}$, $n_c$, and $\Gamma_{12}$. 

Confusion of the conditions of the developed spasing and the spasing threshold is not unusual in the literature -- see, e.g., Ref.\ \onlinecite{Khurgin_Sun_opt_Expr_2012_Impossibility_Electrically_Pumped_Spaser}. Note that the difference between the conditions of the developed spasing and the spasing threshold is numerically much more pronounced for a nanoscopic spaser, which spases at $N_n\gtrsim 1$, than for a macroscopic laser, which lases at $N_n\gg 1$. 

It is also important to investigate what currents are required for the spaser pumping, and whether real nanowires can withstand them. Consider a current required to sustain a single SP per the spasing mode, which can be found from Eq.\ (\ref{N_n}) as $J_{QW}=2e\gamma_n$. This is illustrated in Fig.\ \ref{Spasing_Conditions} (c) for the three plasmonic metals under consideration. 
To compare, a good semiconductor nanowire, such as the channel in a high-performance metal-oxide-semiconductor field-effect transistor (MOSFET) \cite{Packan_et_al_IEDM_2009_32_nm_FET_Technology}, has a $\sim 20-30~\mathrm{nm}$-width and supports working (drive)  current $\sim 20-30~\mathrm{\mu A}$. As comparison with Fig.\ \ref{Spasing_Conditions} (c) shows, such a nanowire is sufficient to pump both  the Ag- or Au-based spasers in their respective plasmonic regions. However, the current required to pump the Al-core nanospasers is large, $J_{QW}\sim 200~\mathrm{\mu A}$. Note that even if such currents are supplied, the Al-based nanospasers may not generate because  the existing gain media main not provide the gain required for the spasing, which is an independent condition  -- cf. Fig.\ \ref{Spasing_Conditions} (b).

Another quantity of interest for the extreme quantum regime is the current that a single-quantum-channel nanowire conducts at the minimum potential drop  required for the one-electron-to-one-SP conversion, $U=\hbar\omega/e$, which is given by Eq.\ (\ref{J_QW}) for $N_c=1$. This is plotted in Fig.\ \ref{Spasing_Conditions} (c) by the dashed line. As one can see that current is rather large: it is sufficient not only for Ag- and Au- but also for Al-core spasers. 


Now we discuss the role of optical phonons, and whether they can dissipate so much energy that the ballistic electron acceleration to the required energy of $\hbar\omega_n$ becomes impossible.  In a semi-classical picture, excitation of optical oscillations with frequency $\omega_{op}$ (emission of an optical phonon) occurs when an electron reaches optical phonon energy $\hbar\omega_{op}$, which requires  acceleration time $t_a=\left(2m^\ast \omega_{op}/\hbar\right)^{1/2}L/\omega_n\approx 10~\mathrm{fs}$, where $m^\ast$ is electron effective mass, the applied electric-field force is $\hbar\omega_n/L$,  $L\sim 100~\mathrm{nm}$ is the nanowire's length, and we adapt values for GaAs: $m^\ast=0.067 m$, and $\hbar\omega_{op}=36~ \mathrm{meV}$. Obviously,  this acceleration time is too short, $t_a\ll t_{op}$, comparing to period of optical phonons $t_{op}=2\pi/\omega_{op}\sim 100~\mathrm{fs}$. Thus this classical picture is inapplicable: on the time scale of $t_a\sim 10~\mathrm{fs}$ the nanowire lattice appears ``frozen'', and the optical phonons cannot manifest themselves.

Instead, one has to invoke quantum mechanics where an electron acquires energy by undergoing a quantum transition with frequency $\omega_n$ rather than the gradual classical acceleration. In Landauer's electron transport picture \cite{Landauer_1985_Conduction_Quantum}, such quantum transitions occur with a period of $2\pi/\omega_n\sim1-3~\mathrm{fs}$ for  $\omega_n$ in the optical range. Thus the emission of optical phonons in our case is  analogous to that following an electronic transition in semiconductor crystals. To take all or a significant part of the electronic transition energy, $\sim\hbar\omega_n\sim 1-3~\mathrm{eV}$, it would require emission of $\sim \omega_n/\omega_{op}\sim 25-80$ optical phonons, whose probability is obviously negligible. 

The electron-optical phonon coupling in direct bandgap semiconductors such as InGaAs is known to be particularly weak \cite{Devreese_2005_Polarons_in_Encylopedia_of_Physics}: Fr\"olich coupling constant $\alpha=0.05-0.06\ll1$ in contrast to ionic solids such as, e.g.,  KCl where $\alpha=3.4$. Consequently, the effects of the electron-phonon coupling may only be significant  for resonant intersubband-type transitions
whose frequencies are in a $30-40~\mathrm{meV}$ range \cite{Lugli_et_al_PRB_1995_Electron_OP_Ibteraction_in_GaAs_QWs} but not for optical-frequency transitions: the emission of optical phonons accompanying a transition in the $\sim 1-3~\mathrm{eV}$  range cannot consume a significant part of the transition energy.

Moreover, in our case there is not only poor temporal matching  ($\omega_n\gg\omega_{op}$) and weak coupling ($\alpha\ll1$) between the electronic transitions and optical phonons but also a very poor spatial overlap: the characteristic optical phonon wavelength, which is on the order of the lattice constant, $\sim a\approx 0.4 ~\mathrm{nm}$, is much less than the extension, $\sim L\sim 100~\mathrm{nm}$, of the electron transition current. This leads to another suppression factor of $\sim (a/L)^2\ll1$. Based on the above arguments, we can safely conclude that the emission of optical phonons under our conditions is unlikely to significantly deplete energy $\hbar\omega_n$ acquired by electrons in the quantum wire.

As the concluding discussion, we have found conditions of the developed CW spasing for the electric pumping in the extreme quantum case, i.e., via a ballistic contact QW with a single conduction quantum, $N_c=1$, under the minimum required potential drop, $U=\hbar\omega_n/e$. It is of fundamental importance that the contact nanowire possesses ballistic conductance to allow for accumulating the maximum electron energy and to prevent heat production in the nanowire and preserve its integrity under the required current load. Note the regime that we have considered for a single conduction quantum, $N_c=1$, is the most stringent: QWs with $N_c>1$ or non-quantum-confined ballistic nanowires will always supply higher currents, relaxing the spasing conditions and allowing for higher SP numbers. As we have shown, the Ag- and Au-core electrically-pumped nanospasers have excellent spasing characteristics for realistic QWs but there are problems for  Al as the spaser-core plasmonic metal. 

Above in this Letter, we have shown a fundamental possibility of the electrically-pumped spaser, or rigorously, that  such a spaser is not fundamentally impossible even for a single-channel QW. We have considered an idealized spaser in the extreme quantum limit but not  problems associated with its experimental realization, including coupling of the QW to the gain medium needed to utilize the kinetic energy accumulated by the electrons upon their passage of the QW. Note that the  ballistic nanowires with necessary characteristics are well studied and widely used, in particular, as channels of the high performance MOSFETs. 

Because currents $J_{QW}\sim 10-100 ~\mathrm{\mu A}$ that we predict to be sufficient for the developed spasing are on the same order as those in the common MOSFETs \cite{Packan_et_al_IEDM_2009_32_nm_FET_Technology}, which also have comparable nanoscopic sizes, we expect that there will be no severe problems with  management of the heat released, which is mostly due to the Ohmic losses in the plasmonic metal. The heat dissipation in the electrically- and optically pumped spasers should be on the same order of magnitude. Note that the optically-pumped strongly 3d-confined nanospasers do work without a damage in the wide range of intensities  \cite{Noginov_et_al_Nature_2009_Spaser_Observation, Gwo_et_al_NanoLett_2011_Plasmonic_Green_Spaser_GaN, Odom_et_al_Nano_Lett_2012_Bowtie_Spaser_Array, Gwo_et_al_Science_2012_Spaser}.

With regard to potential experimental implementations, one of the possibilities is provided by conventional diode gain media used in both spasers (nanolasers) \cite{Hill_et_al_2007_Nat_Phot_2007_Nanolasers, Hill_et_al_Opt_Expr_2009_Polaritonic_Nanolaser, Hill_et_al_Opt_Expr_2011_DFB_SPP_Spaser} and in microlasers \cite{Park_at_al_Nat_Commun_2012_InGaAsP_Microlaser_Graphene_Electrode}. In such a case, the diode's emitting region should be within the plasmonic eigenmode spatial extension. Another possibility for mid-infrared spasers is to use a quantum-cascade gain medium \cite{Capasso_et_al_APL_1995_QCL_Plasmonic_Waveguide} also located within the plasmonic eigenmode extension. In both these cases, the diode or the quantum-cascade wells should be properly biased so that the energy accumulated by an electron as a result of Landauer's transition is matched to the quantum energy of the electron-emitting region. Yet another possibility is the proposed use of the Schottky contact as the gain system \cite{Fedyanin_Opt_Lett_2012_Electrically_Pumped_Spaser} where the semiconductor electrically contacts the plasmonic metal. In such a case,  the kinetic energy of the hot electron is directly transformed into that of the plasmon  \cite{Fedyanin_Opt_Lett_2012_Electrically_Pumped_Spaser} in a process reversed with respect to generation of hot electrons by plasmon decay \cite{Levy_et_al_Nano_Lett_2011_Schottky_Photodetector,
Halas_et_al_Photodetection_with_Plasmonic_Schottky_Nanoantennas}.

Based on this Letter's results, we conclude that achieving an electrically-pumped nanospaser is possible fundamentally and plausible practically. Such a nanospaser will be especially important  for future optoelectronic processors. It will serve as an optical source for SPP-waveguide \cite{Gramotnev_Bozhevolnyi_Nat_Phot_2010_Review_SPPs_Concentration, Han_Bozhevolnyi_Rep_Progr_Phys_2013_SPP_Review} interconnects between transistors, which will eliminate delays and heat production related to the capacitive charging/recharging of the electric interconnects and increase the processing rate to $\sim 100$ GHz transistor-limited speed. In perspective, the nanospaser can also replace the transistor as the logic active element, which will potentially bring about all-optical information processing to $\sim 10-100$ THz  spaser-limited speed.


This work was supported by Grant No. DEFG02-01ER15213 from the Chemical Sciences,
Biosciences and Geosciences Division and by Grant No. DE-FG02-11ER46789 from the Materials Sciences and Engineering Division of the Office of the Basic Energy Sciences, Office of Science, U.S. Department of Energy, and by Chinese Academy of Sciences Visiting Professorship for Senior International Scientists (Grant No: 2012T1G0031) and National Science Foundation of China (Grant No. 61274038). MIS is grateful to V. Apalkov for useful discussions.


\end{document}